# A New Natural Gamma Radiation Measurement System for Marine Sediment and Rock Analysis

M.A. Vasiliev[a)], P. Blum[a)], G. Chubarian[b)], R. Olsen[b)], C. Bennight[a)], T. Cobine[a)], D. Fackler[a)], M. Hastedt[a)], D. Houpt[a)], Z. Mateo[a)], Y. B. Vasilieva[a)]

[a]Integrated Ocean Drilling Program, Texas A&M University, 1000 Discovery Drive, College Station, Texas 77845, USA
[b]Cyclotron Institute, Texas A&M University, College Station, Texas 77843-3366, USA

## Abstract

A new high-efficiency and low-background system for the measurement of natural gamma radioactivity in marine sediment and rock cores retrieved from beneath the seabed was designed, built, and installed on the *JOIDES Resolution* research vessel. The system includes eight large NaI(Tl) detectors that measure adjacent intervals of the core simultaneously, maximizing counting times and minimizing statistical error for the limited measurement times available during drilling expeditions. Effect to background ratio is maximized with passive lead shielding, including both ordinary and low-activity lead. Large-area plastic scintillator active shielding filters background associated with the high-energy part of cosmic radiation. The new system has at least an order of magnitude higher statistical reliability and significantly enhances data quality compared to other offshore natural gamma radiation (NGR) systems designed to measure geological core samples. Reliable correlations and interpretations of cored intervals are possible at rates of a few counts per second.

## 1. Introduction

Natural gamma radiation (NGR) has been studied for the past century as a phenomenon in sediments and rocks, to characterize and interpret geologic strata, for geochemical prospecting of radioactive minerals, and for estimating radiogenic heat production in the Earth (Joly [1], Faul [2], Adams and Gaspirini [3]). Most instruments and procedures to measure NGR in sediments and rocks have been developed by the petroleum industry for borehole logging applications (e.g., Brannon and Osoba [4], Killeen [5], Flanagan et al. [6]). The mining industry developed similar systems for prospecting uranium from airplanes (e.g., Gunn [7], Dickson et al. [8], Grasty et al. [9]).

Lovborg et al. [10] presented the first account of a device and procedures to measure NGR routinely on long sediment cores and perform spectral data analyses. The Integrated Ocean Drilling Program (IODP), and its predecessor, the Ocean Drilling Program (ODP), have been running a shipboard scintillation detector device on the drill ship *JOIDES Resolution* (*JR*) to log core sections drilled from the deep seafloor since 1993 (Blum et al. [11]). ODP and IODP expeditions have recovered, and measured natural gamma radioactivity from more continuous core sections than any other field program over more than two decades. The primary purpose of NGR measurements in scientific ocean drilling is to generate time series of total counts from each core to support stratigraphic correlation between cores from adjacent holes as well as correlation between data obtained from cores and from downhole logging. In addition to aiding depth-correlation, NGR records also provide environmental proxies in pelagic systems, textural and compositional variability on continental margins sedimentary systems, and sequence boundaries in carbonate and clastic systems.

The NGR device used on the *JR* between 1993 and 2003 had design issues that critically limited the scientific value of the measurements. On high-recovery expeditions, up to 8 km of core may be recovered in as little as 40 operational days, which means that each 1.5-m long core section can only be afforded a 10-min residence time at any one core logging station. The old NGR measurement device was mounted along with other types of sensors in a multisensor logger, which reduced the effective NGR counting time to less than 3 min out of the 10 min. Furthermore, the four orthogonally arranged standard 3-in NaI detectors measured only one point in the core section at a time. This reduced the counting time per measurement point for a typical measurement spacing of 5 cm to ~5 s. The majority of sediments studied by ODP/IODP have gamma ray activities that are an order of magnitude lower than those typical for continental crust, often in the range of single-digit counts per second (cps) for typical 3-4-in NaI scintillators. The low count rates obtained that way were associated with large statistical errors. Cosmic and environmental background counts with the old, graded lead-shielded system were on the order of 13 cps. Therefore, the effect to background ratio was often very poor.

Based on this experience, we applied three design principles to the new NGR system to improve performance of the measurement device and the quality of the data. (1) Build a stand-alone system to maximize counting time during the 10-min sample residence time. (2) Build a multi-detector system to count at multiple measurement points simultaneously, thus increasing count rates by an order of magnitude and minimizing statistical error. (3) Drastically reduce background from cosmic rays and environmental radiation by introducing active (plastic scintillators) as well as passive (layers of lead) shielding around the sample chamber.

The opportunity to build the new system came with the large-scale overhaul of the ship and shipboard laboratory as part of the Scientific Ocean Drilling Vessel (SODV) project. The new NGR logger was completed and installed on the renovated *JR* in 2009.

## 2. Instrument Design

### 2.1 Overview

The new NGR system consists of eight NaI detectors. The detector array has passive (layers of lead) and active (plastic scintillators) shielding to reduce background from environmental and cosmic radiation. The entire system is mounted on a sturdy frame with overall dimensions of 212 cm long, 93 cm wide, and 203 cm height, capable of supporting the 5 metric ton total mass. A special design was implemented to achieve mechanical stability under dynamic loading conditions even at high seas. A general view of the NGR system and a cross section of a single shielded NaI detection module are presented in Figures 1 and 2, respectively. Detailed description of system elements will be presented below.

## 2.2 NaI Detector Array

Detector selection and design were defined by the specific requirements of measuring low-activity marine sediments and rocks in plastic liners of nominally 6.6 cm diameter and 0.5-1.6 m length at a rate of up to 10 m/h and measurement spacing of no less than 10 cm, in the shipboard laboratory environment. The detectors had to have sufficient energy resolution to resolve contributions by the isotopes $^{238}U$, $^{232}Th$ series, and $^{40}K$. In order to meet these requirements and collect as high counting statistics as possible, we decided not to collimate gammas from the core (which would enhance space resolution and may be done on shore systems; see Radulescu et al. [12]) but rather use as many large custom-shaped scintillator detectors as can be placed along the core measurement axis considering the system's spatial resolution. Such a configuration provides the highest geometrical efficiency. Taking into account the reliability and maintenance requirements at sea with reasonable cost, we selected NaI(Tl) for our main detectors and plastic scintillators for the cosmic shielding detectors.

The optimal thickness, and therefore total number of NaI detectors that can be placed along the core sample axis, was defined from Monte-Carlo simulations using the GEANT-3 simulation package developed at CERN [13]. The objective was to minimize (<2%) the occurrence of double hits, i.e., detection of signals in two adjacent NaI detectors from the same gamma event. Each detector was to be as big as possible to obtain the highest possible detection and photopeak efficiency. Monte-Carlo simulations obtained for an uncollimated natural gamma source consisting of a mixture of $^{238}U$, $^{232}Th$, and $^{40}K$ isotopes (using the spectral information from Kogan et al. [14]) in a cylindrical geological core (using IODP specifications) yielded a space resolution of 18-20 cm for the proposed system geometry. Minimizing double hits, maximizing space resolution, and achieving reasonably good energy resolution (<10%) determined the maximum size of the NaI crystals.

The final design included eight custom half-ring shaped NaI(Tl) detectors, 4-in (~10 cm) thick both along the core and in direction perpendicular to the core (Figure 3), with 3-in, low-noise, low-K borosilicate glass Philips XP3330 photomultiplier tubes (PMT) as readout devices. The NaI detectors were produced by Saint-Gobain as half-moon shaped crystals packed in stainless steel housings. The energy resolution measured by Saint-Gobain and confirmed by our studies was 7.0%~8.4% on the $^{137}Cs$ 662 keV line. The detectors are placed with distance between centers equal to 20 cm and separated from each other by 7-cm thick, low-background lead spacers. Later measurements on the completed NGR system yielded intrinsic spatial resolution, defined as the full width at half maximum (FWHM) of the response curve, of 16.4 cm for $^{137}Cs$, 16.3 cm for $^{152}Eu$, and 18.1 cm for $^{60}Co$ calibration sources (Figure 4). The NaI crystals have internal impurities from radioactive $^{40}K$ of less than 0.5 ppm, while contamination of $^{238}U$ and $^{232}Th$ isotopes is negligibly small according to Saint-Gobain data. Each NaI detector was equipped with an ORTEC ScintiPack-296 photomultiplier base with preamplifier and high-voltage power supply. Between each photomultiplier and ScintiPack an 8 cm thick low-background lead plug was installed to reduce gamma background from $^{40}K$, which is abundant in ScintiPack internal elements.

## 2.3 Passive Shielding

The NaI scintillators used in our design have worst resolution at low energies ([15] and references therein), and we routinely cull the γ-rays with energy below 100 keV. This eliminates the need to suppress X-rays created by cosmic radiation in the passive shielding, which have energy levels below 100 keV. The best solution for passive shielding is therefore to use pure lead. More complex graded shielding, which typically uses layers of copper and tin in addition to lead, are often selected with high-resolution (even at low energy) Ge detectors in order to suppress undesired X-ray contribution to the low-energy parts of spectra. Such graded shielding is not required for our system due to the applied culling.

The passive shielding designed for this system must suppress gammas from cosmic radiation, for which the plastic scintillators, employed in the active shield (see below), have very low efficiency. Passive shielding must also suppress effectively the soft part of cosmic ray electromagnetic showers generated above the active shielding, which otherwise could significantly increase the counting rates from the plastic scintillators. High counting rates from the plastic scintillators, as will be shown below, would increase random coincidences with γ-rays from the NaI detectors originating in the samples, which would be undesirable. Between the detectors, 70-mm thick low-background lead separators shield the detectors from radiation originating in remote parts of the core samples. A lead thickness of 4 cm above the plastic scintillators also acts as radiator for electromagnetic showers, in which both photons and electron-positron pairs are produced. For high-energy electromagnetic showers the maximum electron-positron pair production in the radiator corresponds to approximately six radiation lengths of material [16], or ~3.6 cm of lead thickness. Thus, 4 cm of lead above the plastic scintillators provides a strong signal in the active shield, as plastics have 100% efficiency for charged particle detection.

Passive shielding also must suppress environmental gamma rays, which in our case may originate mainly from $^{60}Co$ decay, the isotope that usually contaminates at some level the new steel used in vessel laboratory module construction. Monte-Carlo simulations show that a lead shield thickness of 8 cm around each NaI detector will decrease ~1.5 MeV gamma rays generated by the ship's steel about 100 times.

Common virgin lead is internally radioactive in the order of at least 150–250 Bq/kg due to contamination of beta active $^{210}Pb$ isotope [17, 18] and contributes to the gamma background above 100 keV. For the traditional experimental systems, where cosmic radiation dominates the background continuum, the contribution from virgin lead radioactivity can usually be neglected. However, for extremely low background measurements, done either deep underground or, as in our case, using an active shield against cosmic radiation, contributions from internal lead radioactivity become significant and must be reduced as much as possible, particularly for the study of marine core samples that are of relatively small volume and in general have low levels of internal radioactivity. Detailed discussion of lead internal radioactivity may be found elsewhere, for instance see [17, 19] and references therein.

Given the activity of the core sample material, the internal activity of NaI scintillators, and the cosmic radiation that still leaks through the shielding, the tolerance for the lead shield's intrinsic activity was found to be in the order of a few Bq/kg. We use very low activity lead (from Plombum FL, Cracow, Poland) with 4.9 ± 1.2 Bq/kg (measured with Glow Discharge Mass Spectrometer by Dr. D. Leonard at University of Alabama). It is approximately 30-50 times less radioactive than ordinary virgin lead [17]. The previous NGR system on the *JOIDES Resolution* used old lead with 13.7 ± 2.8 Bq/kg (also measured by Dr. D. Leonard). Such low-radioactivity lead is rare and expensive, prompting us to consider a design in which ordinary and low-activity lead are layered at an optimal ratio.

Monte-Carlo simulation using GEANT-3 software shows that 98%-99% of all counts from internal lead radioactivity originate from an area within 2 cm of the detector, the same result also reported in [19]. Therefore, at least 25% of all our lead shielding (about 2 cm around each NaI scintillator) should be low activity. Taking into account technological aspects of lead casting, 4 cm thick internal lead parts were made from low-background lead. To eliminate scattered $\gamma$-rays from the neighboring detectors as well as radiation from the other parts of the core between the detectors, 70-mm thick low-background lead separators were placed between the detectors (see Figure 1). Outside lead parts also have a thickness of 4 cm and are made from virgin Doe Run lead. A total of 1600 kg of low-radioactivity lead was used. The total weight of all lead shielding is about 3.5 tons. We also considered and simulated with GEANT other shielding designs, where the low-background lead layer was replaced with alternate materials, such as tungsten or clean OFHC (Oxygen-Free, High Conductivity) copper. These alternative solutions were finally dropped due to their higher weight and overall manufacturing cost.

## 2.4 Active Shielding

Active shielding is employed to reject signals generated in the NaI(Tl) scintillators from high-energy cosmic radiation, including deeply penetrating muons. Our estimations for cosmic muons showed that active shielding can reduce total background 1.5-2 times for the 0–3 MeV energy region with NaI detectors of general Saint-Gobain purity and surrounded by low-background lead with internal radioactivity at a level of 5 Bq/kg as passive shielding.

The active shield is comprised of five 40 cm wide, 2 cm thick, and 30 cm radius U-shaped plastic scintillators to form a segmented half-cylinder covering NaI detectors and 32-cm radius, flat, half-circle shaped plastic scintillators to cover the front and back openings of the half-cylinder (Figure 5). Each of the seven pieces of plastic scintillator is equipped with two 2-in ETI 9266B photomultiplier (PMT) readouts and P-14 high-voltage divider assemblies supplied by Saint-Gobain.

Background measurements on the completed system, and background reduction due to both passive and active shielding are presented in Figure 6. Measured reduction in background is within the calculated Monte-Carlo range. Active shielding reduces the background by a factor of 1.5–1.8, depending on the internal radioactivity of individual NaI detectors.

## 2.5 Data Acquisition and Control System

Data acquisition and readout systems were assembled from commercially available NIM electronic modules produced by ORTEC, Phillips Scientific, and CAEN. Spectrometric signals from slow outputs of the ScintiPack 296 were amplified with ORTEC 855 Spectroscopy Amplifiers and digitized with ORTEC 927 Multichannel Analyzers (MCA). For rejection of counts in NaI detectors associated with cosmic rays, fast-slow coincidence logic was implemented. In the event of coincidence within a 500 ns window between signals from the fast outputs of NaI detectors and any of the seven plastic scintillators, a VETO signal is generated on the gate input of the MCAs and further readout of such an event is rejected. The status of the gating signal can be defined through Maesto-32 or ScintiVision software commercially available from ORTEC. Correct timing for signals from both NaI and plastic detectors as well as the optimal coincidence time window width were defined empirically through a series of experiments performed with radioactive sources, core samples, and cosmic radiation. Present data acquisition logic is well suited for low counting rates (less than a few thousand events per second per NaI detector), and in our case overall dead time is less than 0.2%.

Routine measurements of natural gamma radiation are operated via the NGR Control software, which provides a user interface that enables visiting scientists and technical support staff to perform consistent and reliable measurements and to store the data for subsequent retrieval, analysis, and interpretation. The measuring software is implemented using LabVIEW, which interfaces with the MCAs through a vendor-supplied driver. The system provides direct access to gamma counts in the MCA buffers. Data are written in standard ASCII format. For each type of experiment, calibration, background survey, and routine sample measurement, the user interface provides spectral plots, real-time summaries of current total counts, and energy spectra for all eight detectors. Users can provide experimental parameters, such as counting time, and set the background and calibration files to be used through the interface. Data reduction is supported through an external library written using C#. The library supports two modes of operation, automatic with pre-configured parameters, or stand-alone, leaving the operator in full control.

## 2.6 Sample Introduction and Motion Control

The sample introduction device is a separate unit. It is designed to deliver 1.5-m long geological core sections placed on a titanium sample holder safely, reliably, and precisely to their measurement positions in the sample chamber. The sample holder rests on a core loading structure in front of the measurement system and is attached to a 55-cm long Delrin rod that is driven by a 2-m stroke NSK ball screw actuator with a 5:1 gearhead reduction powered by a Galil© NEMA 23 brushless motor (Figure 7). The motor is connected to a Galil© PCI bus motion controller (DMC-1842) via an amplifier (AMP-19520) and a 100-pin high-density SCSI cable. The initial sample position was set using an Acuity AR200™ laser displacement sensor, and subsequent position feedback to the Galil controller is given by a 1000-line differential quadrature encoder embedded in the motor. Sample positioning accuracy is within few millimeters and mainly defined by the accuracy of core section placement in sample holder.

The ball screw actuator pushes the sample into the first and then the second measuring positions inside the sample chamber, resulting in an effective measuring spacing of 10 cm. The sample chamber consists of an acrylic pipe with Teflon rods to support the sample holder and rests on top of the NaI detectors and lead separators along the axis of the device. After measurements at both positions are complete, the sample holder is pulled out to the home position and ready for the next sample. Sample introduction is controlled by the same LabVIEW application that provides for data capture and storage.

## 3. System Operation

### 3.1 Energy Calibration and Background Measurements

An energy calibration and low-energy cut off threshold for each NaI detector is performed with a set of standard γ-ray 1-μCi active $^{137}$Cs, $^{60}$Co, and $^{152}$Eu sources. The total energy range is set up to 3000 keV to cover high-energy γ-rays from the $^{232}$Th and $^{238}$U radioactive series.

Counting rates of individual NaI detectors is verified by measuring two synthetic core standards containing $^{40}$K and $^{232}$Th centered above the detectors. The differences between the total counts above the 100 keV thresholds from the eight detectors should be below 1%. For the purpose of the future quantitative spectral analysis, we are using cylindrical standards that represent actual geometrical distribution with known quantity of radioactive sources in the core. Calibration is being performed routinely at least once per month.

The background is measured periodically, ideally whenever the ship's location changes, or at least once per month. A data file can be generated for each NaI detector and measurement position, with the titanium core boat and an empty core liner in place to create conditions as similar as possible to those during subsequent routine measurements. Background measurements are taken for a much longer period of time than for routine sample measurements to minimize contributions to the statistical error. A typical measurement time is 2 x $10^4$ seconds.

With the active shielding on, the background measured using the NaI detectors is on the order of 4 to 8 cps and mainly depends on the amount of $^{40}$K contamination in each individual NaI(Tl) detector and the geographical location of the ship.

### 3.2 Edge Effect Corrections

Given the significant axial length of the detectors' response functions (Figure 4) and the finite axial length of the core samples, edge effects are unavoidable. At measurement position 1, the upper end of the core section is positioned exactly at the center of the last

NaI detector. The lower end of the core section will be positioned at some distance from any of the other NaI detectors, depending on core section lengths. This positioning results in counting rates that are lower than expected on two detectors near both ends of a core section.

The edge effect is expressed as a function of the distance $D$ between a core section's edge and the center of the end detector. For the end detector and position 1, $D = 0$ and the correction factor is 2, i.e., the counting rate is assumed to have half of the magnitude it has when the detector's axial response length is completely covered by the core. If $D < 0$ (less than half of NaI detector is covered by core section) the data are ignored and are not stored or analyzed. For $D > 0$, empirical correction factors were determined at 1-cm intervals using synthetic core standards with $^{40}K$ and $^{232}Th$ isotopes. The correction coefficients represent ratios of the total counts with edge at distance $D$ to the total counts measured with the 47-cm long standard's midpoint centered exactly above the NaI detector (Figure 8).

The measurement results showed that edge correction coefficients need to be applied for 0 cm $< D <$ 18 cm. The results also indicated that the edge correction coefficient is a function of the $\gamma$-ray energy compositions from different isotopes and the positioning accuracy of the standard (Figure 8). The coefficients applied routinely are the average values obtained from $^{40}K$ and $^{232}Th$ standards.

### 3.3 Data Reduction

Data reduction is performed automatically when the user saves a set of 16 core sample data files using the system's user interface. First, the energy calibration coefficients are applied to the sample and background raw spectral data, using the appropriate calibration files. Next, the counts >100 keV are summed for each sample data file as well as the appropriate background file, and the statistical counting errors for each are computed. At this point the edge correction is applied to the sample total counts, altering the total counts but not affecting the statistical error based on lower actual counts. Finally, the background counting rates for each measurement position are subtracted from the total counting rates from the core sample, and the statistical errors are propagated accordingly. The data output for the general user consists of the total count rate with absolute and relative error.

## 4 Data Examples

### 4.1 Performance comparison with old NGR System

To compare the performance of the new NGR system with that of the old system and assess if the goals of this project were achieved, we needed to select a few suitable core samples that had been measured on the old system and were accessible for repeat measurements with the new system. One constraint of our experiment was that core sections were routinely split after they were measured on the ship, so we had only section halves at our disposal. The impact of this situation could essentially be mitigated by increasing (doubling) the counting time accordingly for measurements on the new system. Our selection also had to be limited to core section halves that were still intact, i.e., not destroyed by sub-sampling. We also limited the search to sections stored at the Gulf Coast Core Repository (GCR) at Texas A&M University in College Station, Texas, where the new NGR system was being developed, staged, and tested. We found suitable test material in Cores 202-1239B-39X, 40X, and 41X, representing an interval of ~30 m recovered in the equatorial East Pacific from a depth of ~360-390 m below seafloor [20]

The sample material was measured during ODP Leg 202 for 5 s at 5-cm intervals because that was the optimal configuration for the multisensor track given the operational time constraints (Figure 9a). Background was ~13 cps and background-corrected sample counts ranged from ~4 to 34 cps. The 5-s counting times resulted in large counting errors of 10%-20% for the NGR, making it impossible to discern finer structures in the gamma ray signal (Figure 9a). That error can be reduced by stacking (and smoothing) the data from adjacent measurements. We chose to reduce the depth resolution to 10 cm, a value closer to the intrinsic spatial resolution (FWHM) of ~17 cm for both the old and new NGR systems. The filter we applied passes the counts of every other measurement and adds half of the counts of both adjacent measurements. This method maintains the counts measured and reduces the counting error by a factor of ~1/1.4 (Figure 9b). The error due to the background subtraction is negligible because background is measured over much longer time intervals (20,000 s).

The total measurement time spent on Leg 202 per core section was 150 s (30 positions at 5 s). The fact that the NGR was part of a multisensor track extended the total residence time of a core section at that station to ~10 min, including sample motion and instrument idle time.

The new NGR system is a stand-alone device. If we assume a 10-min sample residence time, we can set the measurement time to the equivalent of 5 min at both measurement positions. Because we measured section halves for this test, the measurement time was doubled to 10 min at each of the two measurement positions. Simultaneous measurements with eight detectors at each of the two positions resulted in dramatically higher count rates with the new NGR system, and errors are reduced to 1.5%-2.5%, despite the smaller effect to background ratio due to measuring section halves instead of whole-round cores (Figure 9c).

### 4.2 High-Fidelity Records at Low Count Rates

One of the main applications of the routine NGR shipboard measurements is to correlate and depth-shift records from cored intervals originating in multiple holes. Other sediment proxies such as magnetic susceptibility or color reflectance are often used for this purpose as well but are sometimes inadequate. NGR has been used successfully for this purpose in the past but only in sediment formations with relatively high radioactivity.

An example is provided here using data from six cores, two each in three adjacent holes drilled during the first expedition with the new system (IODP Expedition 321). The recovered material included pelagic sediments consisting mainly of microfossil ooze with virtually no activity and small amounts of siliciclastic material with moderate activity.

Most NGR variability is in the range of 1-4 cps, and conspicuous peaks have 4-7 cps (Figure 10). The data are presented at the nominal depth intervals reported by the driller and at the core composite depth constructed based on NGR data correlation. The results illustrate that due to the excellent effect to noise ratio and reduced statistical errors the new NGR system produced data that allowed for high-fidelity hole-to-hole correlations even with low count rates. Conventional NGR measurement systems for core samples as well as downhole logging systems have errors of close to 100% at rates of a few counts per second, rendering the data useless. This is the first time that low-count NGR data collected under time constraints can be used as one of the most reliable material proxies for stratigraphic correlation.

The claim that these low-count records represent actual variations in the sediment is not only supported by the correlation of data from multiple holes (Figure 10), but can also be demonstrated with independent data. Figure 11 shows the correlation of NGR data with the corresponding gamma ray attenuation density record, which is based on the transmission of a collimated (3 mm) beam from a $^{137}$Cs source through the core and counting of the attenuated signal with a standard 3-in NaI detector. The attenuated signal is calibrated to bulk density using water and aluminum standards (Blum [21]). The co-variations in the NGR and GRA density records are remarkable (Figure 11) and indicate that increased abundance of a lower density material (clay and silt diluting the carbonate) correlates with higher NGR.

### 4.3 Spectral Resolution

Although spectral analysis is not the subject of this article, the ability to acquire spectra that can be post-processed for elemental abundances of $^{40}$K, $^{238}$U, and $^{232}$Th was a basic requirement. Figure 12 demonstrates that the spectra show sufficient energy resolution to identify the main peaks of the $^{238}$U isotope family above and around 100 keV, as well as the $^{40}$K peak.

As a next stage we plan to implement a full spectral analysis for the NGR system that computes elemental concentrations for potassium, thorium, and uranium based on standard proportions of the isotopes $^{238}$U, $^{232}$Th, and $^{40}$K. This will require analysis of data obtained from different core materials using two fundamental approaches: inversion of data using a calibration matrix obtained from independent, quantitative measurements and/or synthetic core standards, and Monte-Carlo simulations of the spectra. Standard data reduction software can then be implemented to provide $^{40}$K, $^{238}$U, and $^{232}$Th data routinely and in real time to the user. An example of a Monte-Carlo simulated spectrum vs. experimentally measured spectrum dominated by $^{238}$U is shown in Figure 12.

## 5. Conclusions

A new natural gamma radiation detection system for the measurement of geological cores aboard the research drill ship *JOIDES Resolution* was constructed that meets the requirements imposed by limited counting times available during high-recovery expeditions. The new system implements a background suppression capability, including passive lead and active plastic scintillator shields, that delivers unprecedented data quality per counting time unit. The often mission-critical procedure of correlating cores from adjacent holes can now take full advantage of the geological gamma radiation profile, even in very low activity Earth materials such as carbonate-rich sediments. The correlation of high-quality NGR data with other physical property records offers new research opportunities. The results achieved with the new system meet or exceed the quality of downhole logging data and thus support the important objective of core-log correlation. The energy resolution of the new system allows determination of $^{238}U$, $^{232}Th$, and $^{40}K$ abundance estimates based on spectral analysis, which will be fully exploited in a future study.

## Acknowledgements

This research was sponsored by the U.S. National Science Foundation (NSF) contract OCE-0352500. We thank Tony Montallano and Ken Greer for their custom machining work; many of our IODP colleagues, the crew and scientific staff aboard the *JOIDES Resolution* for logistical support and operational assistance; and Jeff Fox and Jack Baldauf for their unwavering support of the project.

## Figure Captions

Figure 1. General view of NGR system. The top lead shield is removed to provide a view of the stainless steel covered active shield plates (also see Figures 2 and 5). A central cut-out exposes three of the custom NaI detectors with the axial sample chamber directly on top (also see Figure 3). The hoist mounted onto the steel frame facilitates assembly and removal of shield segments.

Figure 2. Detector-shield assembly, including inner lead and outer lead shields and active shield mounted between the two.

Figure 3. Custom NaI scintillators are half-ring shaped to optimize efficiency of core measurements. The crystals are 4-in thick and 4-in wide, are enclosed in stainless steel housings, and are equipped with PMT at their bases. The detectors are spaced at 20-cm intervals along the measurement axis.

Figure 4. NGR internal space resolution defined as FWHM from measurements with $^{137}$Cs, $^{60}$Co, and $^{152}$Eu calibration sources. Experimental data (symbols) are shown together with Gaussian fit (line).

Figure 5. Geometry and arrangement of plastic scintillators for active shield.

Figure 6. Effect of passive and active shielding in NGR system, as measured on shore before deployment on the ship.

Figure 7. Completed NGR system on the *JOIDES Resolution* research vessel.

Figure 8. Edge effect correction coefficients based on measurements of 47-cm long core standards with $^{40}$K and $^{232}$Th isotopes, respectively.

Figure 9. Comparison of total counts data from drill cores (ODP Hole 202-1239B, Cores 39X-41X) measured with the old NGR system and the new system presented here. A. Whole-round sections measured every 5 cm for 5 s during ODP Leg 202. The large counting error of 14% due to the short counting times was typical during an expedition. B. The data are stacked into 10-cm intervals, which decreases the counting error to ~10%, brings the interval closer to the intrinsic depth resolution of the system (FWHM~15 cm), and makes the sample interval compatible with that of the new system. C. Section halves measured with the new system at the core repository, years after the expedition. Much longer counting times possible with the multisensor system during the same residence time of a core in the lab as in A and B, and improved detector properties, reduce the error to 1.5%.

Figure 10. Example application of NGR data from to depth shifting and stratigraphic correlation, using cores from three holes (U1337A, B, and D) from IODP Expedition 321. A. Core data measured against the original drilling depth have large absolute depth errors due to tides and heave. As shown with the horizontal bars representing core lengths, the soft-sediment cores overlap as a result of gas and elastic expansion during recovery. B. Cores are depth-shifted using NGR data with the goal to maximize cross-hole correlation. As shown with the horizontal bars representing core length, coring gaps exist between cores in any one hole. For example, gaps between Cores 9 and 10 in Hole U1337B and U1337D are covered by Core 10 in Hole U1337A, allowing construction of a complete stratigraphic section.

Figure 11. Example data from Expedition 321 showing how NGR total counts correlate inversely with bulk density data obtained with a gamma ray attenuation (GRA) densitometer. Data points for both data sets are represented as vertical error bars calculated from count rates. Upper curve: NGR; lower curve: GRA. A. Correlation at a scale of tens of meters. Thick lines are curve fits to emphasize the trend. B. Correlation at a scale of meters. Fine lines connect data points in NGR curve.

Figure 12. Comparison of Monte-Carlo simulated spectrum with real data from a marine core. A. Linear counts scale with isotope identification for major peaks. All peaks except for the $^{40}K$ peak belong to the $^{238}U$ series, revealing a uranium-dominated natural gamma ray signature for this sediment. B. Logarithmic counts scale to emphasize lower count (higher energy) part of spectrum.

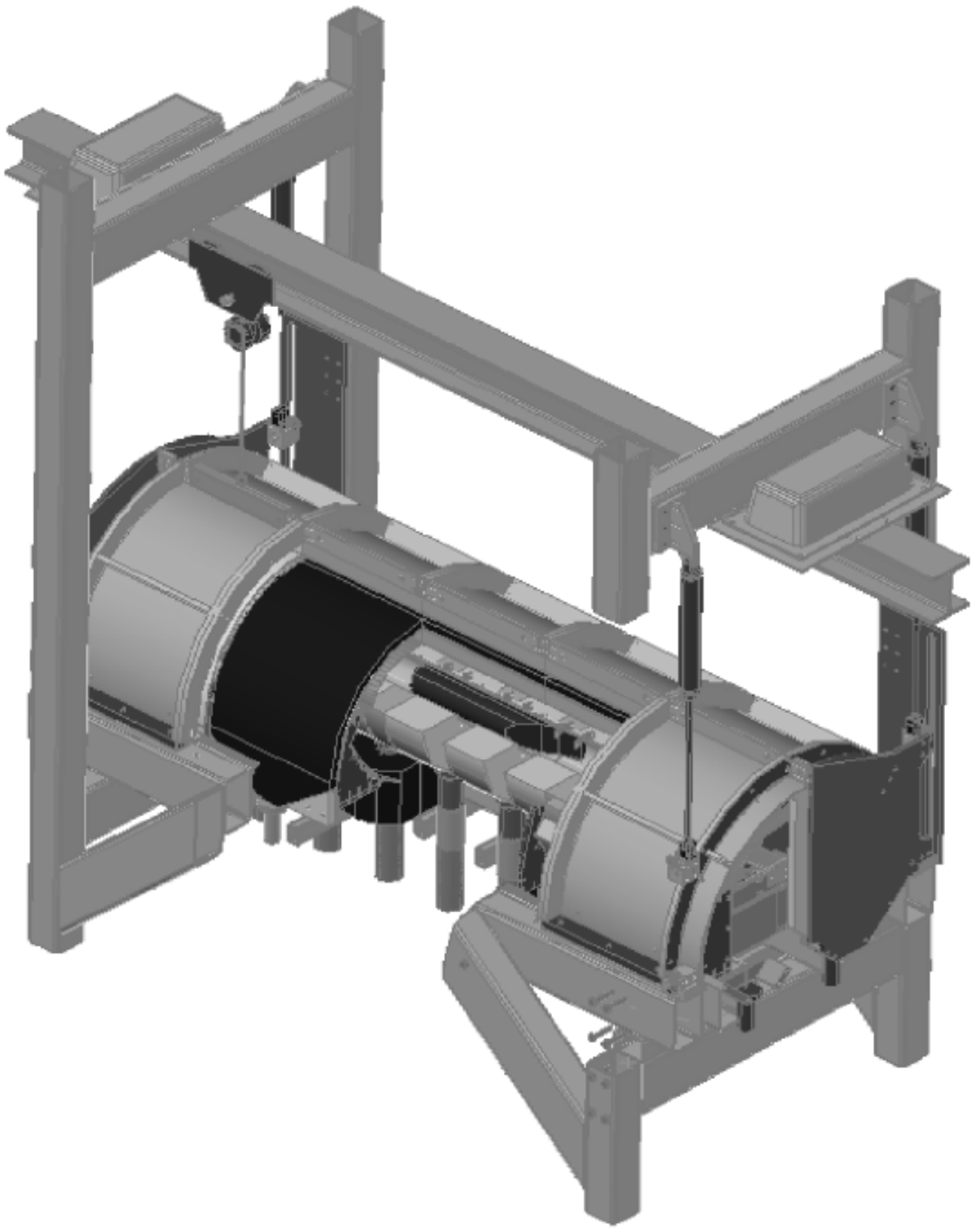

Figure 1 (Vasilyev et al)

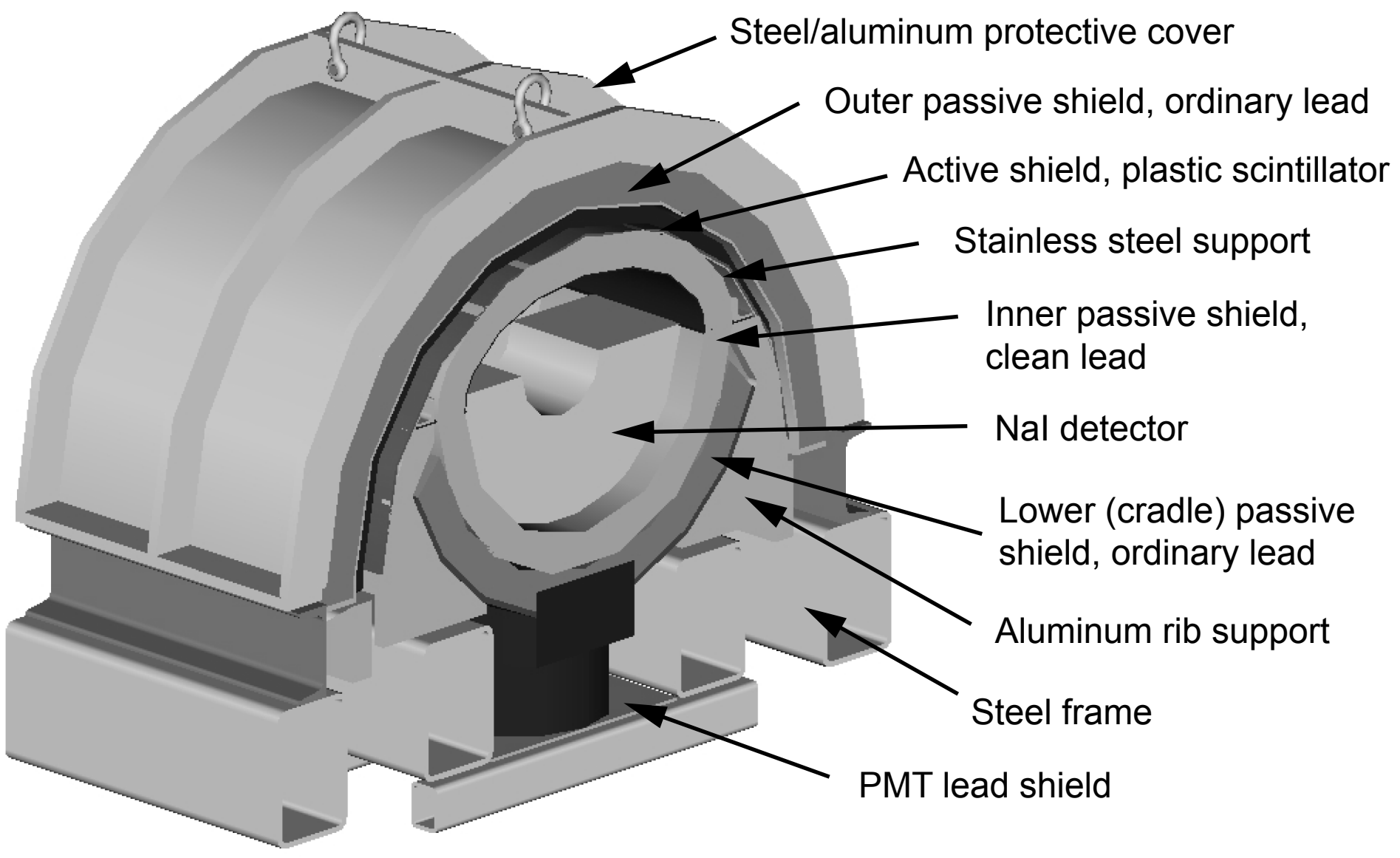

Figure 2 (Vasilyev et al)

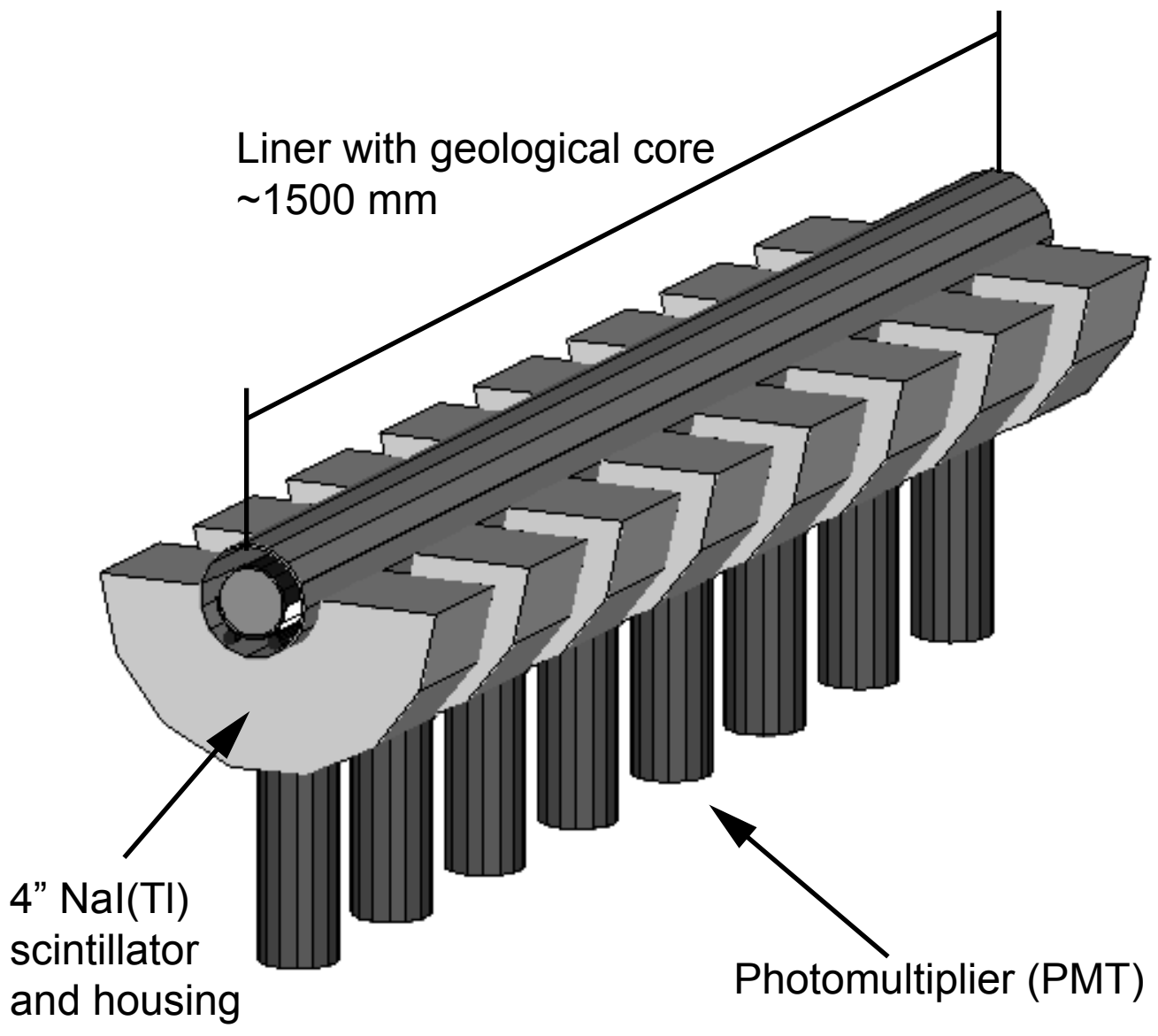

Figure 3 (Vasilyev et al)

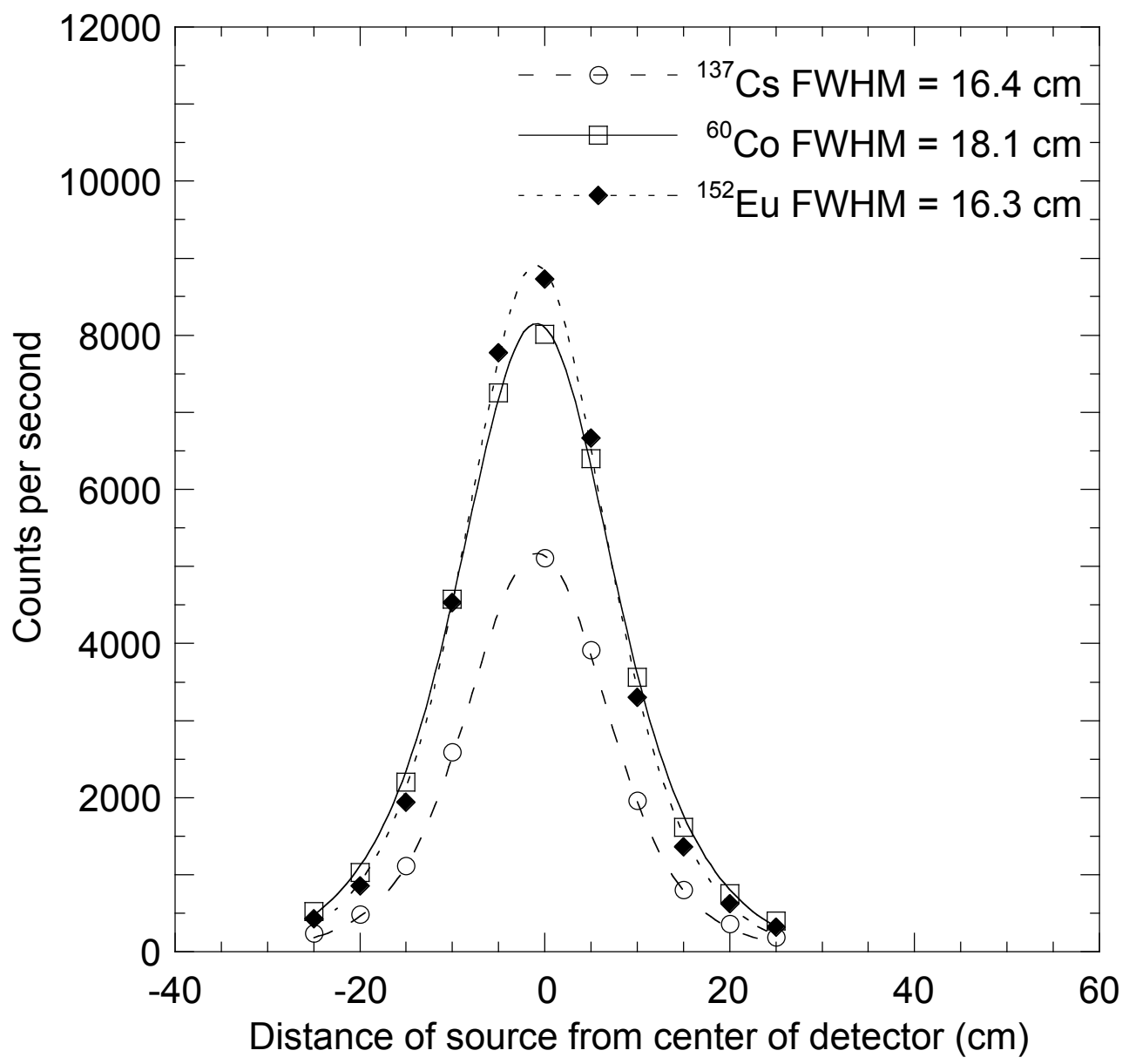

Figure 4 (Vasilyev et al)

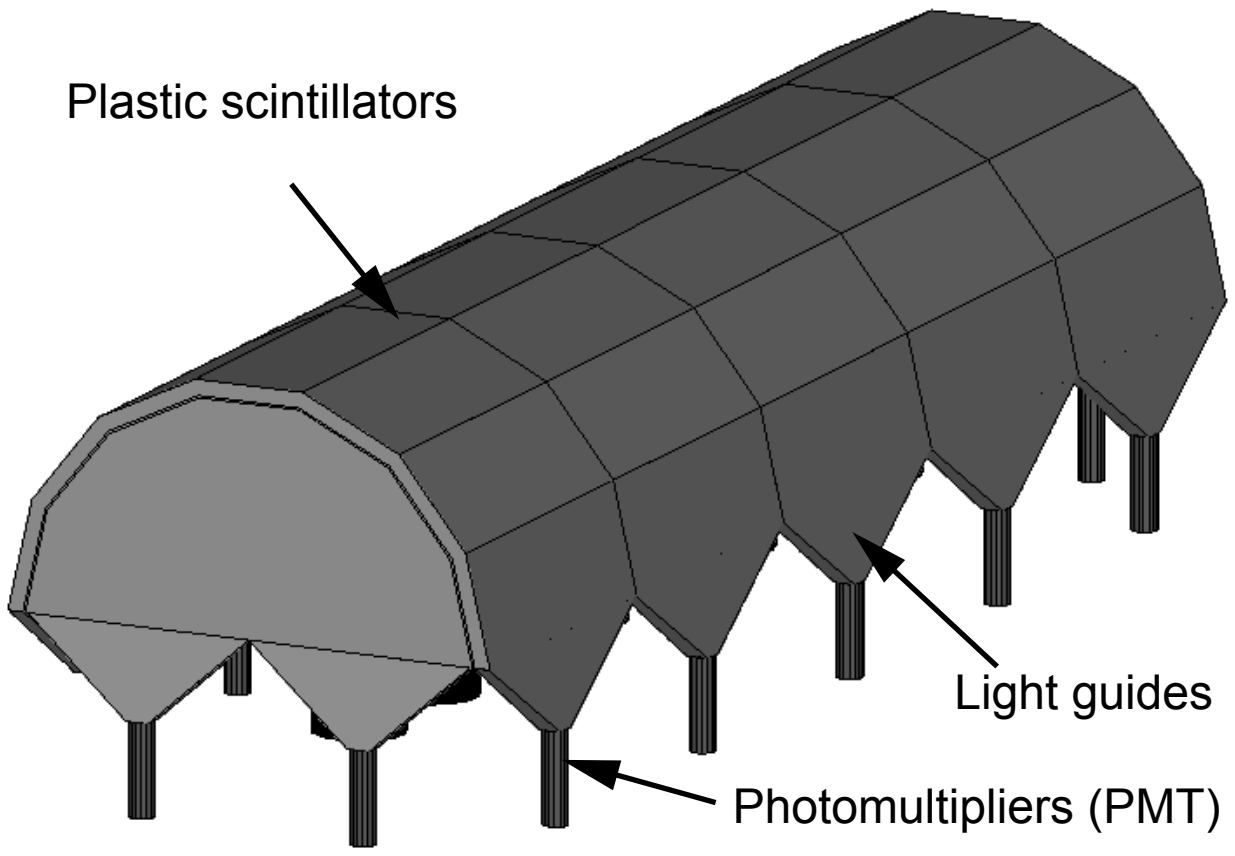

Figure 5 (Vasilyev et al)

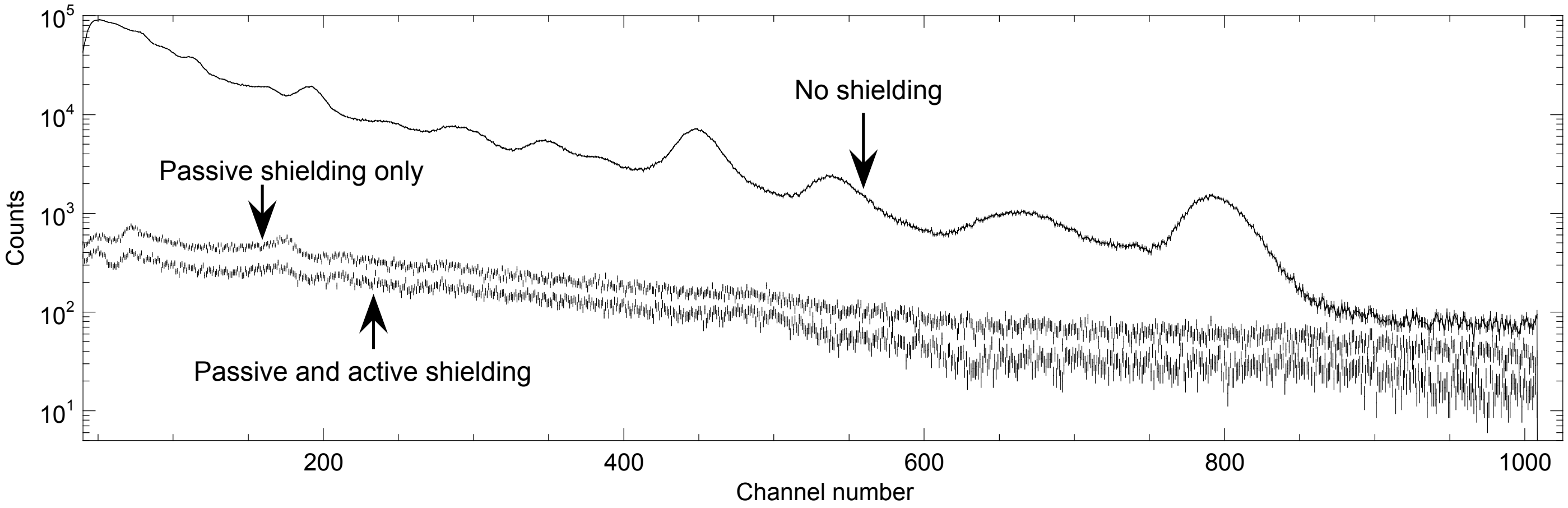

Figure 6 (Vasilyev et al)

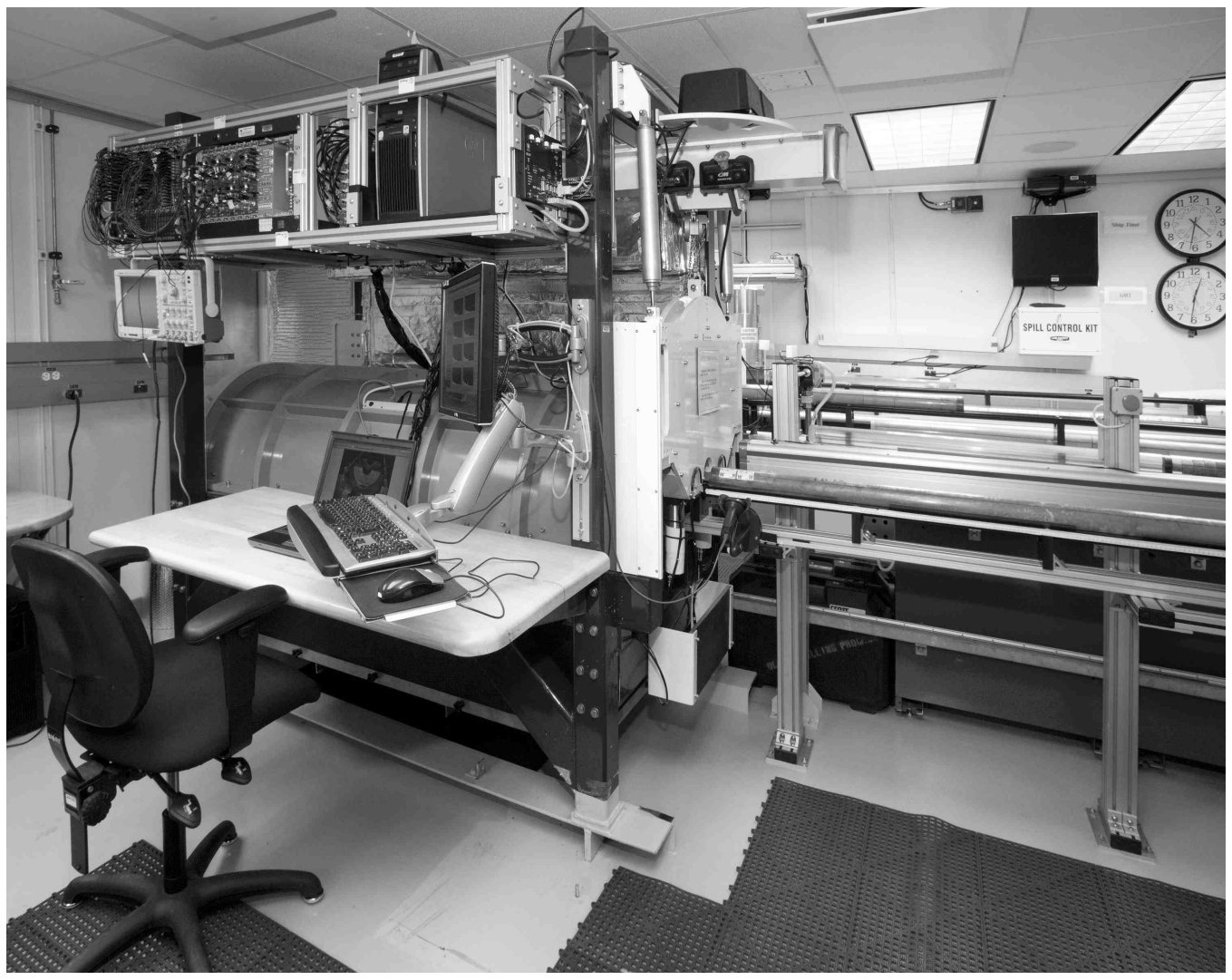

Figure 7 (Vasilyev et al)

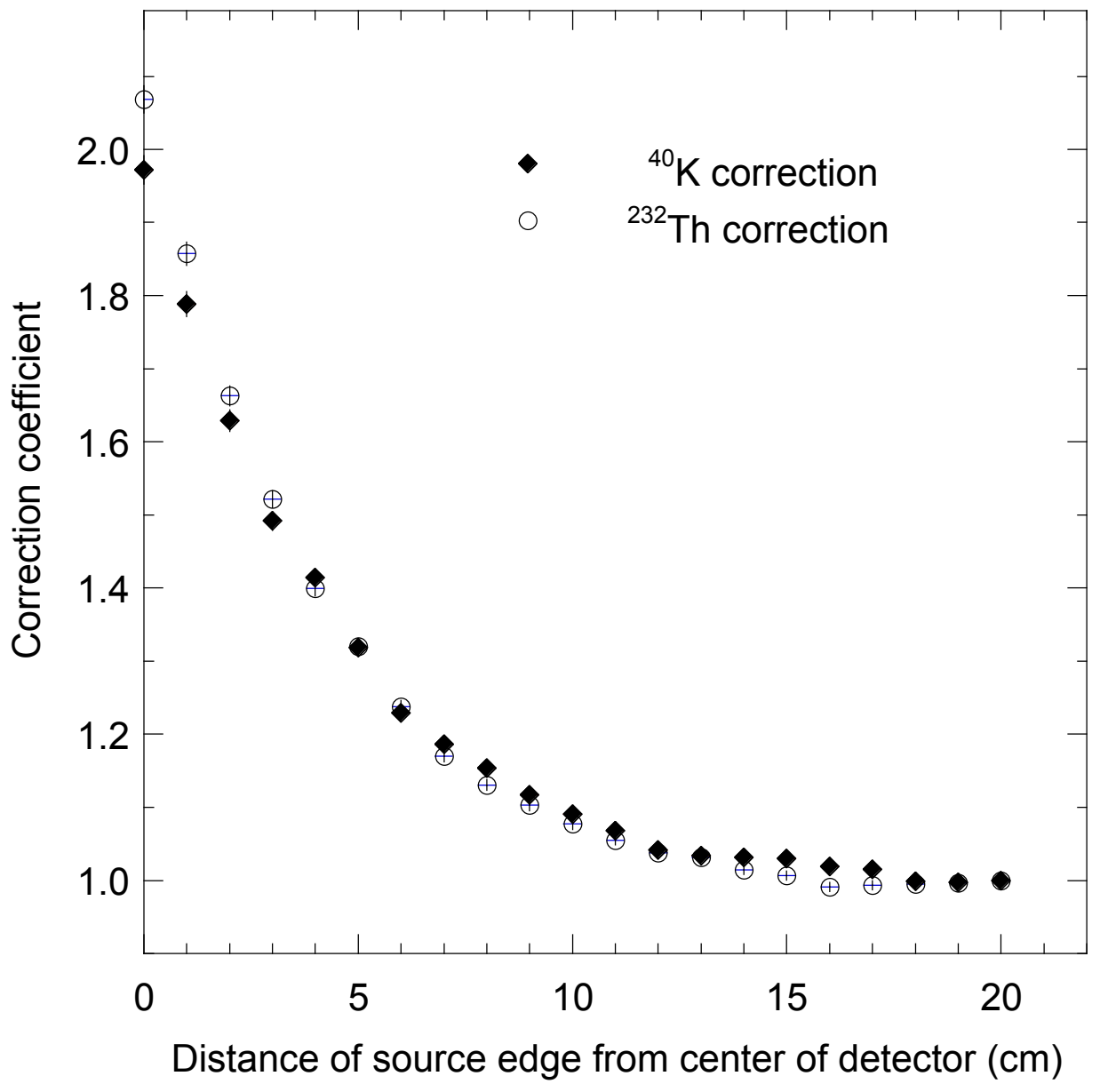

Figure 8 (Vasilyev et al)

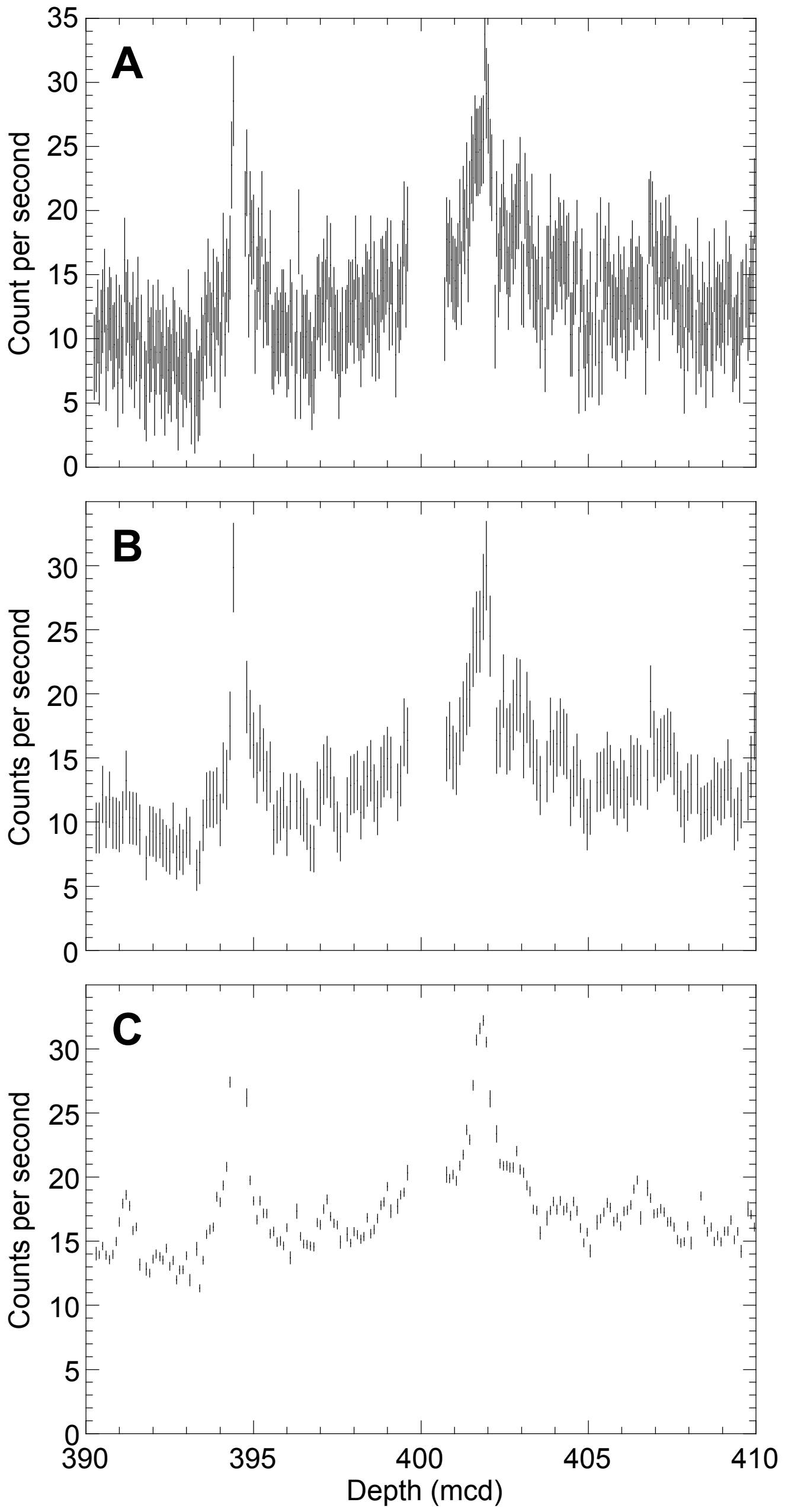

Figure 9 (Vasilyev et al)

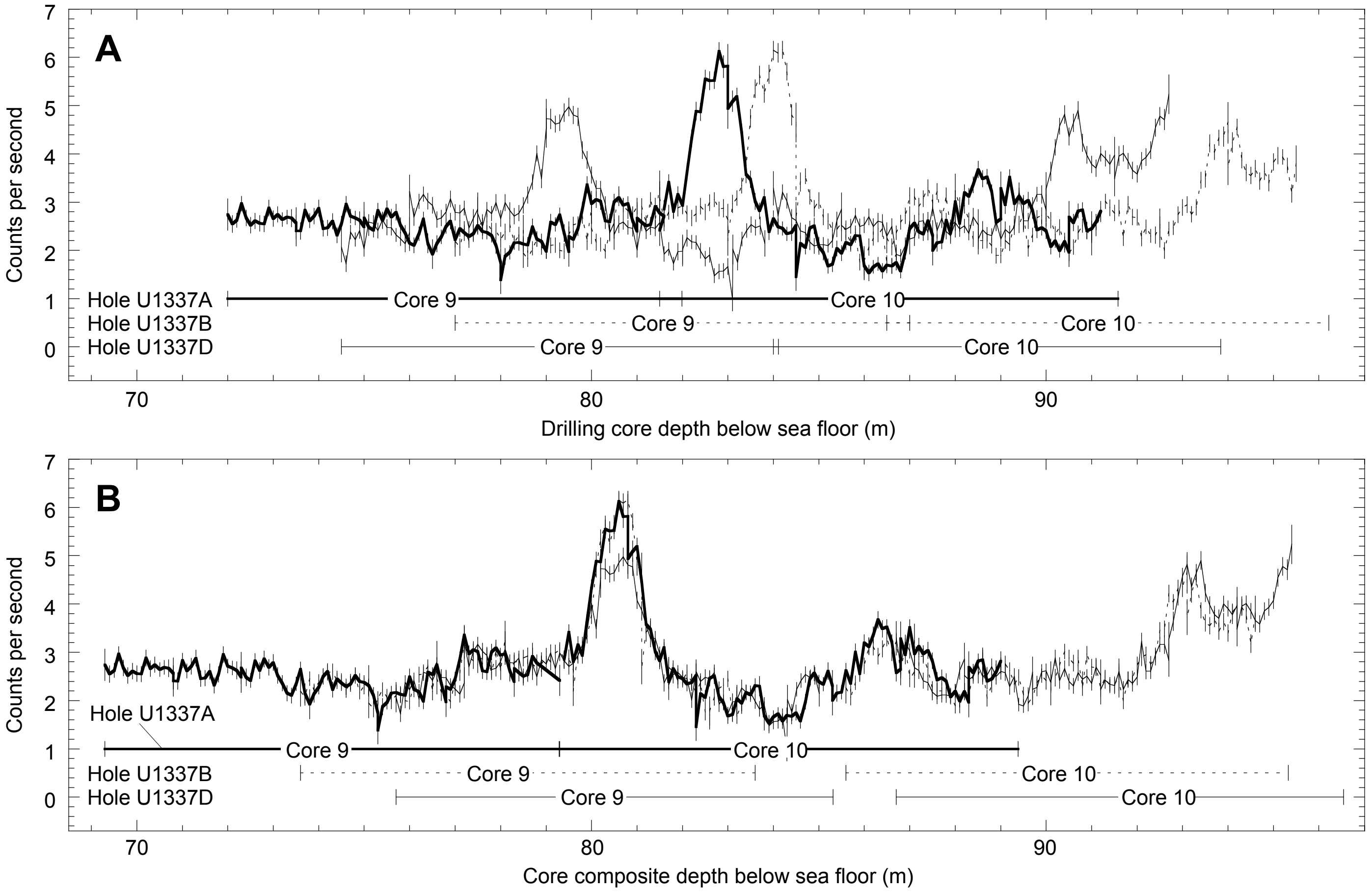

Figure 10 (Vasilyev et al)

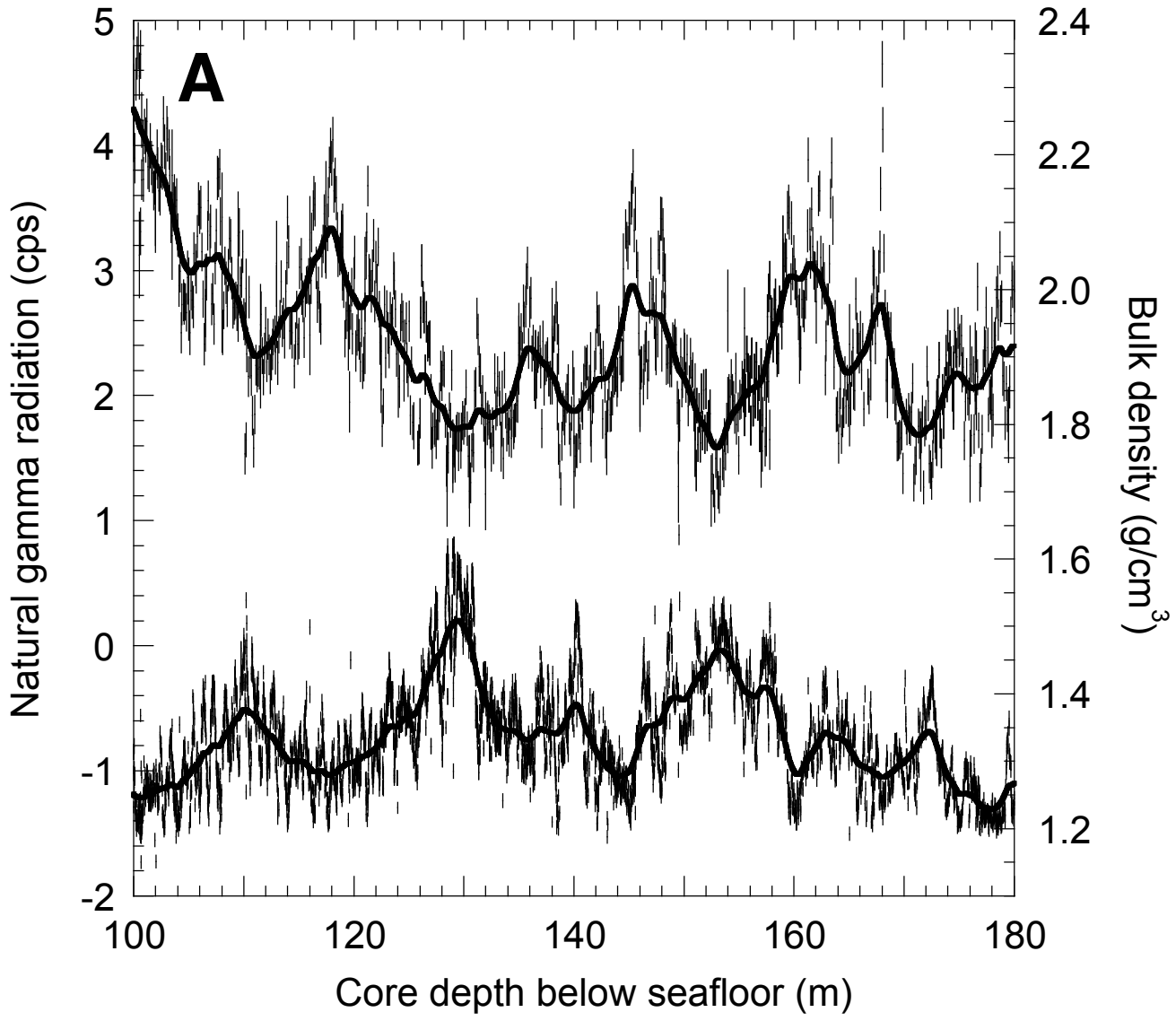

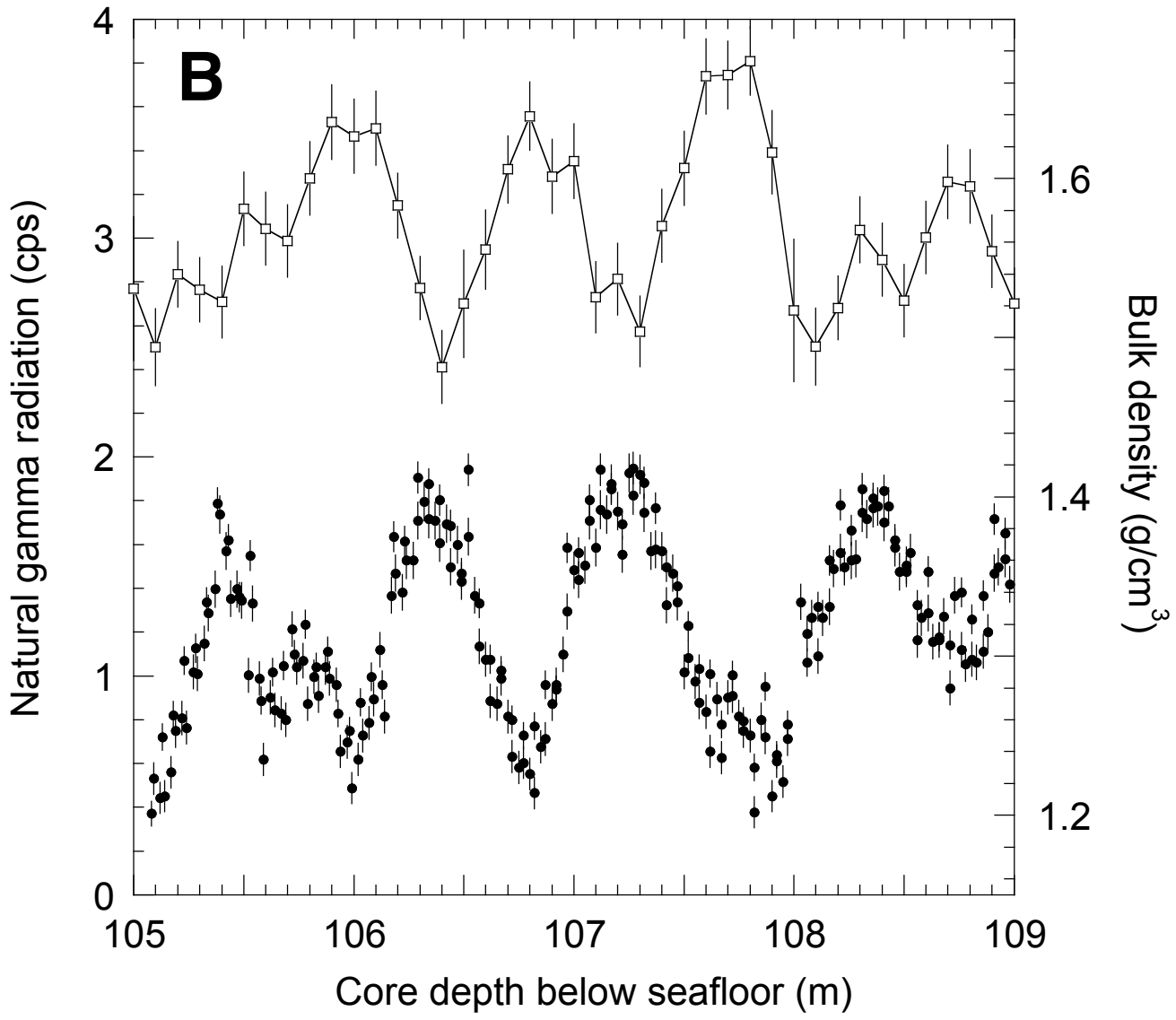

Figure 11 (Vasilyev et al)

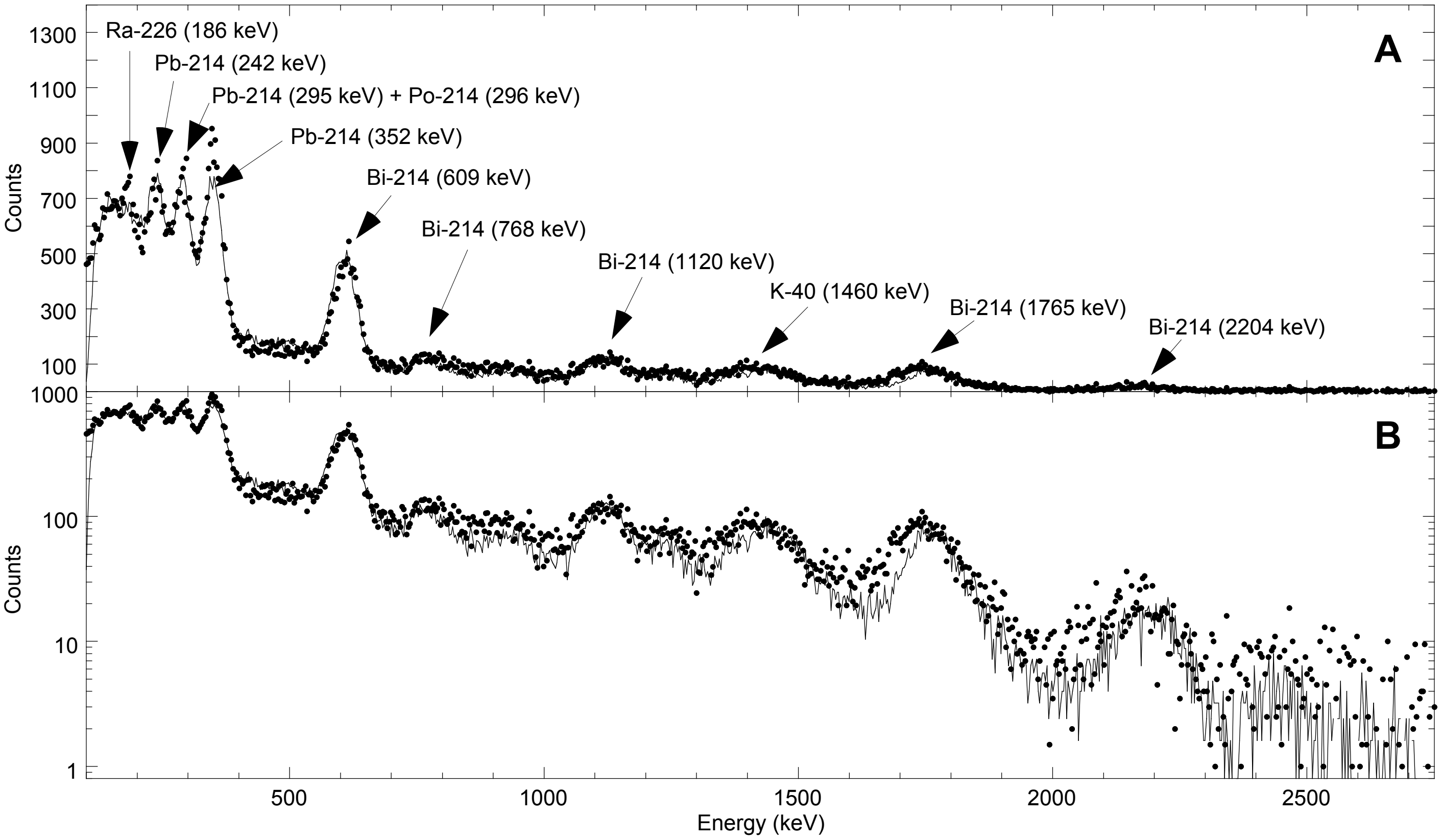

Figure 12 (Vasilyev et al)